\documentclass[aps,prc,twocolumn,amscd,amsmath,amssymb,verbatim,showpacs,notitlepage]{revtex4-1}
\usepackage{hyperref}
\usepackage{graphicx} 
\usepackage{epstopdf} 
\usepackage{subfigure}
\usepackage{multirow}
\usepackage{xcolor}
\usepackage[normalem]{ulem}
\usepackage{array}

\newcommand{\commentout}[1]{}

\begin{document}

\title{Neutrino-nucleus coherent scattering as a probe of neutron density
distributions}
\author{Kelly Patton$^{1}$} \email{kmpatton@ncsu.edu} 
\author{Jonathan Engel$^{2}$} \email{engelj@physics.unc.edu}
\author{Gail C. McLaughlin$^{1}$} \email{gail$\_$mclaughlin@ncsu.edu}
\author{Nicolas Schunck$^{3}$} \email{schunck1@llnl.gov}
\affiliation{$^{1}$Physics Department, North Carolina State University, Raleigh, North Carolina 27695, USA\\
$^{2}$Department of Physics and Astronomy, University of North Carolina, Chapel
Hill, North Carolina 27599, USA\\ 
$^{3}$Physics Division, Lawrence Livermore Laboratory, Livermore, California 94551 USA}
\date{\today}
\begin{abstract}
Neutrino-nucleus coherent elastic scattering provides a theoretically appealing
way to measure the neutron part of nuclear form factors.  Using an expansion of
form factors into moments, we show that neutrinos from stopped pions can probe
not only the second moment of the form factor (the neutron radius) but also the
fourth moment.  Using simple Monte Carlo techniques for argon, germanium, and
xenon detectors of 3.5 tonnes, 1.5 tonnes, and 300 kg, respectively, we show
that the neutron radii can be found with an uncertainty of a few percent when
near a neutrino flux of $3\times10^{7}$ neutrinos/cm$^{2}$/s.  If the
normalization of the neutrino flux is known independently, one can determine
the moments accurately enough to discriminate among the predictions of various
nuclear energy functionals.  
\end{abstract}
\maketitle 
          
%
%

\section{Introduction}

The size of a nucleus is one of its most fundamental properties.  Although the
distributions of protons in nuclei are well known, the neutron distributions
are comparatively poorly constrained.  A precise measurement of neutron radii
could have important implications in both nuclear physics and astrophysics.

In nuclear physics the most common framework for predicting neutron densities
is energy density functional theory (DFT) --- Skyrme or relativistic --- with
parameters that are at least in part fitted to other nuclear observables or
pseudo-data such as nuclear-matter properties, root-mean-square radii, atomic
masses, etc. \cite{Bender:2003166}. Recent efforts to determine new generations
of energy density functionals have pointed to the complex correlations among
the values of physical quantities that they predict \cite{UNEDF0,UNEDF1}.
Among observables used to optimize functionals, the neutron form factor is
particularly important because it determines the neutron skin and radii, which
in turn are strongly correlated in density functionals with the symmetry energy
and incompressibility of nuclear matter \cite{Reinhard:2010}. Precise
measurements of neutron radii could, therefore, significantly improve the
predictive power of energy functionals.

In astrophysics, the size of the neutron skin may have important implications
for neutron stars. While there are accurate measurements of pulsar orbital
periods and masses, the radius of neutron stars, as well as their moments of
inertia or gravitational redshift remain poorly constrained by observation, and
must be provided by theory \cite{Fattoyev:2010, Li:2006, Steiner:2010,
Steiner:2012}.  Such global properties as masses, radii, and composition, are
determined by the equation of state (EOS) of neutron-rich nuclear matter, which
involves such quantities as the density dependence of the nuclear symmetry
energy.  As just noted, the symmetry energy is itself strongly correlated with
the neutron skin \cite{Steiner:2005}. Precise measurements on the neutron skin,
therefore, provide information about the equation of state of neutron matter
and thus the size of neutron stars \cite{Horowitz:1999fk}.

Traditional measurements of neutron radii have involved hadronic scattering,
with typical uncertainties of order 10\%.  Parity-violating electron
scattering, used by the PREX experiment at Jefferson Laboratory to measure the
radius of lead, is cleaner.  The parity-violating asymmetry, i.e.\ the
fractional difference in cross section between positive- and negative-helicity
electrons, is roughly proportional to the weak form factor, which is the
Fourier transform of the weak charge density.  If one can measure the asymmetry
(at a single Q$^2$) to 3\% then one can determine the root mean square neutron
radius to 1\%.  The uncertainty on the neutron radius from PREX is about
$2.5\%$ \cite{PhysRevLett.108.112502} but $\pm $1\% may be possible in the
future.

The use of neutrino-nucleus coherent scattering to probe the weak form factor
was first proposed in Ref.\ \cite{Amanik:2009zz}.  The authors considered a one
tonne $^{40}$Ar detector, with a nucleus described by a simplified form
factor.  Their analysis suggested that ton-scale detectors could replicate the
10\% uncertainty of hadronic scattering methods when used in conjunction with a
source of Michel-spectrum neutrinos.  While large detectors are required since
neutrinos interact only weakly, the theoretical interpretation of the results
is straightforward and model independent.  Neutrino-nucleus coherent scattering
has been proposed for a number of other purposes as well, for example to detect
supernova neutrinos \cite{Horowitz:2003cz}, to measure the Weinberg angle
\cite{Scholberg:2005qs,Bueno:2006yq}, to look for a neutrino magnetic moment,
and to search for sterile neutrinos \cite{Formaggio:2011jt}. In all these
cases, a weak nuclear form factor must be either measured or assumed before
useful information can be extracted.

Both neutron and proton distributions in the nucleus affect neutrino-nucleus
coherent scattering, but the neutron distribution has much more leverage.  In
this paper we suggest the use of a Taylor expansion to write the
nuclear-neutron form factor in terms of moments of the neutron density
distribution.  Using this expansion and a simple Monte Carlo simulation, we
show that neutrino-nucleus coherent scattering can probe not only neutron
radii, but also the higher-order moments of neutron distributions. We use the
examples of argon \cite{Scholberg:2005qs}, germanium \cite{Anderson:2011bi} and
xenon targets to show the expected ranges of sensitivity.

The paper is organized as follows. In section \ref{sec:theory}, we introduce
the model used to estimate neutrino-nucleus scattering count rates, including
in our discussion the Taylor expansion of the neutron form factor and the
calculation of the moments of the neutron distribution in nuclear DFT.  In
section \ref{sec:simulations}, we present and discuss the results of the
Monte-Carlo simulations.

%
%

\section{Coherent Scattering and the Form Factor}
\label{sec:theory}

We present in this section the details of the model, including the kinematics
of neutrino-nucleus coherent scattering, the dependence of the neutron form
factor on the moments of the neutron distributions, and the DFT-based
calculations of the moments.

\subsection{Kinematics}
To calculate the cross section for neutrino-nucleus coherent elastic
scattering, we sum the contributions of each nucleon to the amplitude, which we
then square and sum over available phase space. The resulting cross section,
for spherical nuclei (neglecting small corrections from various sources) is
\cite{Barranco:2005}
\begin{multline}
\frac{d\sigma}{dT}(E,T) = \frac{G_{F}^{2}}{2\pi}M \left[ 2 - \frac{2T}{E} +
\left(\frac{T}{E}\right)^{2} - \frac{MT}{E^{2}}\right] \\
 \times \frac{Q_{W}^{2}}{4}F^{2}(Q^{2})\,,
\label{eq:dsigmadT}
\end{multline}
where $E$ is the energy of the incoming neutrino, $T$ is the nuclear recoil
energy, $M$ is the mass of the nucleus, $G_{F}$ is the Fermi constant, and
$Q_{W} = N - (1 - 4 \sin^{2}{\theta_{W}})Z$ is the weak charge of the nucleus
($\sin^{2}{\theta_{W}} \approx 0.231$). The cross section also contains the
form factor $F^{2}(Q^{2})$, which is a function of the momentum transfer
$(Q^{2} = 2E^{2}TM/(E^{2} - ET))$.  One of the neglected corrections is from
higher multipoles in odd-A nuclei, which include the effects of deformation as
well as nonzero spin.  The higher multipoles add terms to Eq.\
(\ref{eq:dsigmadT}) only at order $Q^4$, and even those changes are much
smaller than $\mathcal{O}(1/A)$ for the nuclei considered here.

The form factor corrects for scattering that is not completely coherent at
higher energies. It encodes information about the nuclear densities through a
Fourier transform, which in spherical nuclei takes the approximate form
\cite{Horowitz:2003cz}
\begin{multline}
F(Q^{2}) = \frac{1}{Q_{W}} \int \left[ \rho_{n}(r) - (1 -
4\sin^{2}{\theta_{W}}) \rho_{p}(r) \right] \\
\times \frac{\sin{(Qr)}}{Qr} r^{2} dr \,, 
\label{eq:formFactorIntegral}
\end{multline}
where $\rho_{n,p}(r)$ are the neutron and proton densities.  We have neglected
effects due to the finite size of the nucleons, which alter the relation
between the point-neutron density and the cross section by modifying the form
factor at high $Q$.  These effects could easily be included and would barely
change the results of our sensitivity analysis below. 

The separation of the neutron and proton terms in Eq.\
(\ref{eq:formFactorIntegral}) makes it possible to write the form factor as
\begin{equation}
F(Q^{2}) =\frac{1}{Q_{W}}\left[ F_{n}(Q^{2}) - (1 -
4\sin^{2}{\theta_{W}})F_{p}(Q^{2}) \right] \,.
\label{eq:formFactorPplusN}
\end{equation}
Since the coefficient of the proton form factor is small, the scattering
depends mainly on the neutrons, making neutrino-nucleus coherent scattering
well suited to measuring the neutron distribution. 

There are two primary types of neutrino sources to consider: neutrinos
generated from fission processes in nuclear reactors, and neutrinos from the
decay of stopped pions. Reactor neutrinos have lower energy, resulting in
correspondingly low nuclear-recoil energies. Because background can obscure
low-energy recoil, we consider neutrinos produced from the decay of stopped
pions. Stopped pions are produced in large quantities at both spallation
sources and accelerator sources.  An example of a spallation source is the Spallation Neutron Source
at Oak Ridge National Laboratory, which hits a mercury target with a beam of
protons.  Pions are produced, with negative pions captured in the target and
positive pions coming to rest and decaying. The pions decay through $\pi^{+}
\rightarrow \nu_{\mu} + \mu^{+}$. The muon neutrinos are monoenergetic with an
energy of 29.9 MeV.  The muons then come to rest and further decay via $\mu^{+}
\rightarrow e^{+} + \nu_{e} + \overline{\nu}_{\mu}$. The probability that
neutrinos $\nu_{e}$ or antineutrino $\overline{\nu}_{\mu}$ are emitted in the
range $(E, E+dE)$ read 
\begin{eqnarray}
\label{eq:spectra}
f_{\nu_{e}} & =  & \frac{96}{m_{\mu}^{4}}(m_{\mu}E_{\nu_{e}}^{2} -
2E_{\nu_{e}}^{3})dE_{\nu_{e}}, \nonumber \\ 
f_{\overline{\nu}_{\mu}} & = & \frac{16}{m_{\mu}^{4}}(3
m_{\mu}E_{\overline{\nu}_{\mu}}^{2} -4
E_{\overline{\nu}_{\mu}}^{3})dE_{\overline{\nu}_{\mu}} \,, 
\end{eqnarray}
where $m_{\mu}$ is the mass of the muon. The energy of the neutrinos range up
to $\sim 52$ MeV, which results in typical nuclear recoil energies on the order
of tens of keV to 100 keV. The momentum transfer associated with these energies
runs up to $\sim 100$ MeV. 

To calculate the number of scattering events as a function of recoil energy, we
fold the neutrino spectra with the cross section: 
\begin{equation}
\frac{dN}{dT}(T) = N_{t} C \int_{E_{\rm min}(T)}^{m_{\mu}/2}{ f(E)
\frac{d\sigma}{dT}(E,T) dE }\,, 
\label{eq:dNdT}
\end{equation}
where $N_{t}$ is the number of targets in the detector, $C$ is the flux of
neutrinos of a given flavor arriving at the detector, the normalized energy
spectra $f(E)$ includes all three types of neutrino produced in pion decay, and
$E_{\rm min}(T) = \frac{1}{2}(T + \sqrt{T^{2} + 2 T M})$ is the minimum energy
a neutrino must have to cause a nuclear recoil at energy $T$. The upper bound
of $m_{\mu}/2$ is the maximum energy for a neutrino produced from muon decay at
rest.  

\subsection{Form-factor expansion}
\begin{figure}[hbt]
\includegraphics[width=\linewidth]{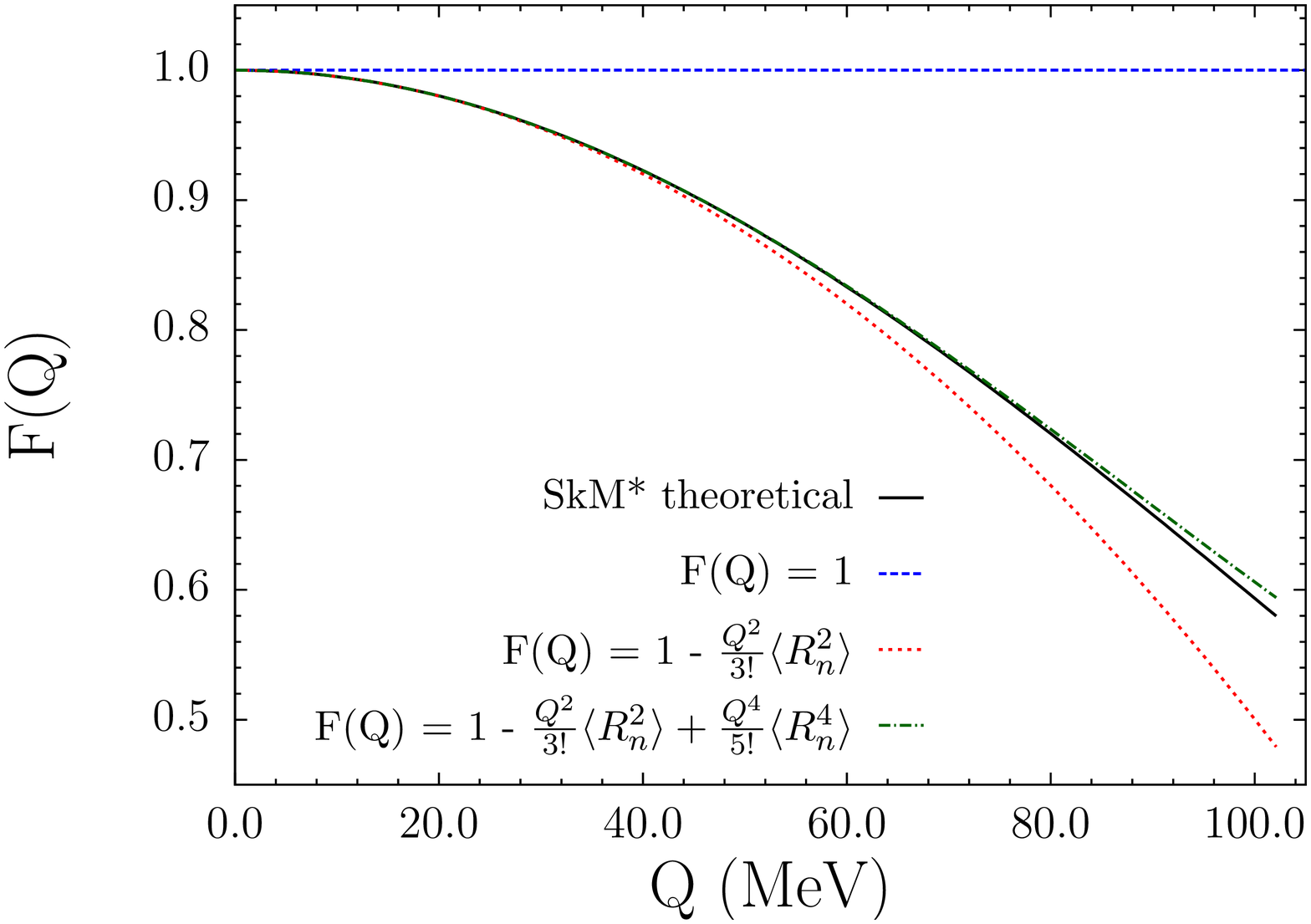}
\caption{(Color online.) The factor $F(Q^2)$ for $^{40}$Ar predicted by the
Skyrme functional SkM$^*$ (solid black curve), and truncations of the expanded
form factor at various orders of $Q$: $Q^0$ (dashed blue curve), $Q^2$ (dashed
red curve), $Q^4$ (solid green curve).  Terminating the expansion at $Q^4$
(with coefficient $\frac{1}{5!}\langle R^{4}_{n} \rangle$) gives good agreement
with the full form factor over the range of $Q^2$ relevant for the scattering
of neutrinos from stopped pion beams.
\label{fig:FFexpansion}}
\end{figure}

Since the form factor is included in the calculation of the number of events,
nuclei with different density distributions will produce different
recoil-energy distributions. The recoil distributions therefore provide a good
test for models that predict the density. We can increase the usefulness of the
recoil distribution by expanding the form factor in $Q$. The dominant neutron
piece can be represented as
\begin{eqnarray}
F_{n}(Q^{2}) & \approx & \int{ \! \rho_{n}(r) \left( 1 - \frac{Q^{2}}{3!} r^{2} 
+ \frac{Q^{4}}{5!}r^{4} - \frac{Q^{6}}{7!}r^{6}  + \cdots \right) r^{2} dr } 
\nonumber \\
 & \approx & N \left( 1 - \frac{Q^{2}}{3!} \langle R^{2}_{n} \rangle +
 \frac{Q^{4}}{5!}\langle R^{4}_{n}\rangle -  \frac{Q^{6}}{7!}\langle R^{6}_{n}\rangle 
 + \cdots \right)\,,
 \label{eq:formFactorExpanded}
\end{eqnarray}
with
\begin{equation}
\langle R_{n}^{k} \rangle  =  \frac{\displaystyle\int{\rho_{n} r^{k} d^{3}r}}{\displaystyle\int{\rho_{n} d^{3}r}}  \,.
\label{eq:momentDefinition}
\end{equation}
Written this way, the form factor is a sum of the even moments of the neutron
density distribution. These moments are straightforward to calculate from the
density, and represent physically relevant and measurable quantities. Since the
neutrinos we consider have relatively low energy, we can truncate the expansion
after just two terms for lighter nuclei such as argon and germanium, and three
terms for heavier nuclei like xenon. As an illustration, we show in Fig.
\ref{fig:FFexpansion} the theoretical neutron form factor predicted by the
Skyrme functional SkM$^*$ \cite{Bar82} for $^{40}$Ar. Including moments up to
$\langle R^{4}_{n}\rangle$ is sufficient to reproduce the full form factor
curve over the relevant range of $Q$ values. In other words, we can fit
experimental scattering data in $^{40}$Ar with just two parameters, $\langle
R_{n}^{2}\rangle$ and $\langle R_{n}^{4}\rangle$. 

\begin{figure}[hbt]
\includegraphics[width=\linewidth]{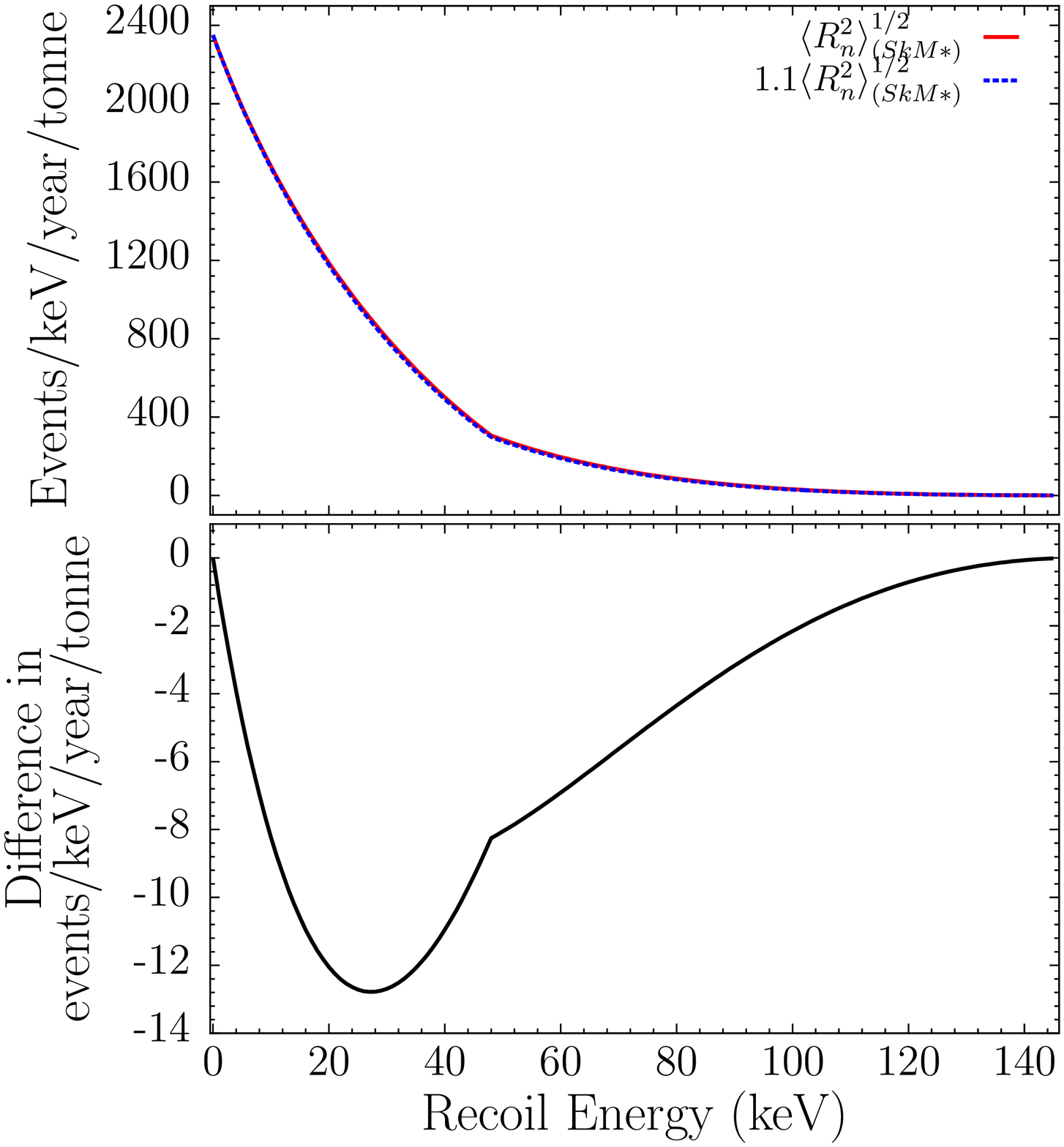}
\caption{(Color online.) Top: Event rates in $^{40}$Ar as a function of recoil
energy, with two different RMS neutron radii. The red (solid) curve represents
predictions of the Skyrme functional SkM$^*$, while the blue (dotted) curve
represents the same for an RMS radius made $10\%$ larger, as described in the
text. The flux at the detector is taken to be $3\times 10^{7}$
neutrinos/cm$^{2}$/sec per flavor. Bottom: The difference between the two
curves on top.}
\label{fig:eventsAndDiffsAr}
\end{figure}

Fig.\ \ref{fig:eventsAndDiffsAr} shows the effects on event rates in $^{40}$Ar
of changing a single important measure of the density distribution, the
root-mean-square (RMS) neutron radius $\langle R_n^2 \rangle^\frac{1}{2}$.  We
produced the figure as follows: First, we calculated event rates as a function
of recoil energy with the functional SkM$^*$, via the expansion in Eq.\
(\ref{eq:formFactorExpanded}).  Next, we did the same with the RMS neutron
radius 10\% larger and no changes to the other terms in the expansion.  The
resulting curves for $^{40}$Ar, with an assumed neutrino flux of
$3\times 10^{7}$ neutrinos/cm$^{2}$/ sec per flavor, appear in the upper panel
of Fig.\ \ref{fig:eventsAndDiffsAr}, while the difference between them is
plotted in the bottom panel.  A 10\% difference in the RMS neutron radius
results in a difference of $\sim$780 events/tonne/year, which is about 1.2\% of
the total event rate over the entire energy range.  This difference is
concentrated at a nuclear recoil energy of 30 keV.  

We obtain similar results in Ge and Xe, though there we must use effective
moments, which we define in the next section, to average over different
isotopes. In Ge, a 10\% difference in the effective RMS radius yields an
integrated difference of $\sim$8100 events/tonne/year, or about 6\% of the 
total, concentrated around a nuclear recoil of 15 keV. In Xe the same change 
in effective RMS radius results in a difference of $\sim$20200 events/tonne/year, 
or about 8\% of the total, and is concentrated at a nuclear recoil energy of 8 keV.

As $Q^{2}$ goes to zero, the form factor approaches $F(0) = 1$, so the low
energy portions of the event-rate curves converge and all the difference curves
go to zero. At high energies, there are very few events, so both event-rate
curves approach zero and the differences also approach zero. The larger
elements have smaller total recoil-energy ranges. In all three cases, however,
the largest difference in the event-rate curves occurs at a recoil energy of
about 20\% of the maximum. We note that the highest and lowest energies will most 
likely be excluded from an experimental analysis because of background at the
detector, and that the effect of changing $\langle R_n^2 \rangle ^{\frac{1}{2}}$ 
is most prominent in the energy range that is experimentally accessible to 
cryogenic detectors such as CLEAN \cite{McKinsey2005355} and CLEAR
\cite{Scholberg:2005qs, CLEAR:2009}, even though more events occur at inaccessibly low
recoil. 

\begin{table}[tdp]
\caption{Isotopes and abundances used in the calculations for germanium and 
xenon \cite{NNDC}.}
\begin{center}
\begin{tabular}{|cc|cc|}
\hline
\hline
Isotope & Abundance & Isotope & Abundance\\
\hline
$^{70}$Ge & 0.205 & $^{128}$Xe & 0.0191 \\
$^{72}$Ge & 0.274 & $^{129}$Xe & 0.264\\
$^{73}$Ge & 0.078 & $^{130}$Xe & 0.041\\
$^{74}$Ge & 0.365 & $^{131}$Xe & 0.212\\
$^{76}$Ge & 0.078 & $^{132}$Xe & 0.269\\
 & & $^{134}$Xe & 0.104 \\
 & & $^{136}$Xe & 0.089 \\
 \hline
 \hline
\end{tabular}
\end{center}
\label{table:isotopes}
\end{table}

\subsection{Effective Moments}
Naturally occurring argon is made at 99.6\% of $^{40}$Ar: the expansion in Eq.\
(\ref{eq:formFactorExpanded}) is therefore sufficient to accurately compute
neutrino-nucleus scattering. Germanium and xenon, however, both have multiple 
isotopes that occur naturally with relatively high abundance (see table 
\ref{table:isotopes}). To account for that fact, we define effective moments 
for these elements as follows: 

To calculate the event curves for germanium and xenon, it is necessary to sum
over all isotopes. There are several isotope-specific quantities in the cross
section, including the mass, neutron number, and the moments of the neutron and
proton distributions. In addition, each isotope will have a different number of
atoms in the detector. To calculate the rate for all isotopes, we therefore use
\begin{eqnarray}
\label{eq:allIsotopesSum}
\frac{dN}{dT}(T) &=& N_{A} M_{\textrm{detector}} C \\
&& \times \int{f(E) \sum_{i}\left[
\frac{X_{i}}{M_{i}}\left(\frac{d\sigma}{dT}(T,E)\right)_{i} \right] dE },
\nonumber
\end{eqnarray}
where the coefficient $N_{t}$ in Eq. (\ref{eq:dNdT}) is now replaced by
$N_{A}M_{\textrm{detector}}$ and a summation over isotopes $i$ of the
cross-section with the weights $X_{i}/M_{i}$. Here, $X_{i}$ is the natural
abundance of isotope $i$, $M_{i}$ is the mass of that isotope, $N_{A}$ is
Avogadro's number, and $M_{\text{detector}}$ is the total mass of the element
(including all its isotopes) in the detector. 

Since the form factor appears squared in the cross section, the sum in Eq.
(\ref{eq:allIsotopesSum}) will have neutron terms, proton terms, and terms that
include both neutron and proton moments (see Eq. (\ref{eq:formFactorPplusN})).
The definitions for the proton effective moments will follow the same pattern
as those for the neutrons, so we will concentrate only on the neutrons at this
point.  

Some algebra and the definitions of the cross section and form factor allow us
to write the sum in Eq. (\ref{eq:allIsotopesSum}) as
\begin{widetext}
\begin{eqnarray}
\sum_{i}\left[ \frac{X_{i}}{M_{i}}\left(\frac{d\sigma}{dT}(T,E)\right)_{i} \right] & = & \frac{G_{F}^{2}}{8\pi}   \left[ \sum_{i} \left( X_{i} N_{i}^{2} \right) \left( 2 - \frac{2T}{E} + \left(\frac{T}{E}\right)^{2}\right) - \sum_{i}\left( X_{i} N_{i}^{2} M_{i} \right) \left( \frac{T}{E^{2}}  \right) \right.   \nonumber \\
 & &  -  \sum_{i} \left( X_{i} N_{i}^{2} M_{i} \langle R_{n}^{2} \rangle_{i}\right) \left( 2 - \frac{2T}{E} + \left(\frac{T}{E}\right)^{2}\right) \frac{Q^{2}}{3 \langle M \rangle} 
 + \sum_{i} \left( X_{i} N_{i}^{2} M_{i}^{2} \langle R_{n}^{2} \rangle_{i}\right) \left( \frac{T}{E^{2}}  \right) \frac{Q^{2}}{3 \langle M \rangle} \nonumber \\
 & &  + \sum_{i} \left( X_{i} N_{i}^{2} M_{i}^{2} \langle R_{n}^{2}
 \rangle^{2}_{i}\right) \left( 2 - \frac{2T}{E} +
 \left(\frac{T}{E}\right)^{2}\right) \frac{Q^{4}}{36 \langle M \rangle^{2}}
 - \sum_{i} \left( X_{i} N_{i}^{2} M_{i}^{3} \langle R_{n}^{2} \rangle^{2}_{i}\right) \left( \frac{T}{E^{2}}  \right) \frac{Q^{4}}{36 \langle M \rangle^{2}} \nonumber \\
 & &  + \sum_{i} \left( X_{i} N_{i}^{2} M_{i}^{2} \langle R_{n}^{4}
 \rangle_{i}\right) \left( 2 - \frac{2T}{E} +
 \left(\frac{T}{E}\right)^{2}\right) \frac{Q^{4}}{60 \langle M \rangle^{2}}
  - \sum_{i} \left( X_{i} N_{i}^{2} M_{i}^{3} \langle R_{n}^{4}
 \rangle_{i}\right) \left( \frac{T}{E^{2}}  \right) \frac{Q^{4}}{60 \langle M
 \rangle^{2}} \nonumber \\
  & & - \sum_{i} \left( X_{i} N_{i}^{2} M_{i}^{3} \langle R_{n}^{2}
 \rangle_{i}\langle R_{n}^{4}
 \rangle_{i}\right) \left( 2 - \frac{2T}{E} + \left(\frac{T}{E}\right)^{2}\right) \frac{Q^{6}}{360 \langle M
 \rangle^{3}} \nonumber \\
   & & \left.- \sum_{i} \left( X_{i} N_{i}^{2} M_{i}^{4} \langle R_{n}^{2}
 \rangle_{i}\langle R_{n}^{4}
 \rangle_{i}\right) \left( \frac{T}{E^{2}}  \right) \frac{Q^{6}}{360 \langle M
 \rangle^{3}}   + \cdots \right] \,,
 \label{eq:complicated-expression}
\end{eqnarray}
\end{widetext}
where $\langle M \rangle = \sum_{i} X_{i} M_{i}$ and 
$Q^{2} = 2 E^{2} T \langle M \rangle/(E^{2} - E T)$.
In the above equation, we have kept all terms that have 
$\langle R^{2}_{n} \rangle$, $\langle R^{4}_{n} \rangle$, or both. (For xenon, 
we use all additional terms that include $\langle R^{6}_{n} \rangle.$) In this 
expression several effective (isotope-weighted) moments occur, two of each order. 
The two effective second moments, after normalization, are 
\begin{eqnarray}
\langle R^{2}_{n} \rangle_\text{eff,1} & = & \frac{\sum_{i}{X_{i}N_{i}^{2} M_{i} \langle R^{2}_{n} \rangle_{i}}}{\sum_{i}{X_{i} N_{i}^{2}M_{i}}}, \label{r21} \\
\langle R^{2}_{n} \rangle_\text{eff,2} & = & \frac{\sum_{i}{X_{i}N_{i}^{2} M_{i}^{2} \langle R^{2}_{n} \rangle_{i}}}{\sum_{i}{X_{}{i} N_{i}^{2}M_{i}^{2}}}, \label{r22}
\end{eqnarray}
and the two effective fourth moments are 
\begin{eqnarray}
\langle R_{n}^{4} \rangle_\text{eff,1} = \frac{\sum_{i} X_{i} N_{i}^{2} M_{i}^{2} \langle R_{n}^{4} \rangle_{i}}{\sum_{i}{X_{i}N_{i}^{2} M_{i}^{2}}}, \label{r41} \\
\langle R_{n}^{4} \rangle_\text{eff,2} = \frac{\sum_{i} X_{i} N_{i}^{2} M_{i}^{3} \langle R_{n}^{4} \rangle_{i}}{\sum_{i}{X_{i}N_{i}^{2} M_{i}^{3}}}. \label{r42}
\end{eqnarray}

The differences between the values of the two moments of each order is small,
as can be seen for the Skyrme functional SkM$^*$ in table
\ref{table:effMomComparison}. In fact, these differences are smaller than the 
typical numerical uncertainties in DFT calculations coming from the truncation 
of the basis or the size of the mesh. We therefore make the approximation 
that the two second moments are equal (calling them 
$\langle R_{n}^{2}\rangle_\text{eff}$) and that the two fourth moments are equal 
(calling them $\langle R_{n}^{4} \rangle_\text{eff}$.) There are also terms in 
the cross section that involve $\langle R_{n}^{2} \rangle^{2}$, from which we 
can define $(\langle R_{n}^{2} \rangle^{2})_\text{eff}$.  Although mathematically 
$(\langle R_{n}^{2} \rangle^{2})_\text{eff}$ is not equal to 
$(\langle R_{n}^{2} \rangle_\text{eff})^{2}$, numerically they are very similar, 
as shown in table \ref{table:effMomComparison}. After equating them and making 
the similar approximations just described, we can rewrite 
Eq.\ \ref{eq:complicated-expression} in terms of effective moments as
\begin{widetext}
\begin{eqnarray}
\sum_{i}\left[ \frac{X_{i}}{M_{i}}\left(\frac{d\sigma}{dT}(T,E)\right)_{i} \right] & = & \frac{G_{F}^{2}}{8\pi}   \left[ \sum_{i} \left( X_{i} N_{i}^{2} \right) \left( 2 - \frac{2T}{E} + \left(\frac{T}{E}\right)^{2}\right) - \sum_{i}\left( X_{i} N_{i}^{2} M_{i} \right) \left( \frac{T}{E^{2}}  \right) \right.   \nonumber \\
& & - \langle R_{n}^{2}\rangle_\text{eff} \sum_{i} \left( X_{i} N_{i}^{2} M_{i}\right) \left( 2 - \frac{2T}{E} + \left(\frac{T}{E}\right)^{2}\right) \frac{Q^{2}}{3 \langle M \rangle} 
 + \langle R_{n}^{2}\rangle_\text{eff} \sum_{i}\left( X_{i} N_{i}^{2} M_{i}^{2}
\right) \left( \frac{T}{E^{2}}  \right) \frac{Q^{2}}{3 \langle M \rangle}
\nonumber \\
&& + \langle R_{n}^{2}\rangle_\text{eff}^{2} \sum_{i} \left( X_{i} N_{i}^{2} M_{i}^{2}\right) \left( 2 - \frac{2T}{E} + \left(\frac{T}{E}\right)^{2}\right) \frac{Q^{4}}{36 \langle M \rangle^{2}} 
 - \langle R_{n}^{2}\rangle_\text{eff}^{2} \sum_{i}\left( X_{i} N_{i}^{2} M_{i}^{3} \right) \left( \frac{T}{E^{2}}  \right) \frac{Q^{4}}{36 \langle M \rangle^{2}} 
\nonumber \\
&& + \langle R_{n}^{4}\rangle_\text{eff} \sum_{i} \left( X_{i} N_{i}^{2}
M_{i}^{2}\right) \left( 2 - \frac{2T}{E} + \left(\frac{T}{E}\right)^{2}\right)
\frac{Q^{4}}{60 \langle M \rangle^{2}} 
- \langle R_{n}^{4}\rangle_\text{eff} \sum_{i}\left( X_{i} N_{i}^{2} M_{i}^{3}
\right) \left( \frac{T}{E^{2}}  \right) \frac{Q^{4}}{60 \langle M \rangle^{2}}
\nonumber \\
&& - \langle R_{n}^{2}\rangle_\text{eff} \langle R_{n}^{4}\rangle_\text{eff} \sum_{i}
\left( X_{i} N_{i}^{2} M_{i}^{3}\right) \left( 2 - \frac{2T}{E} +
\left(\frac{T}{E}\right)^{2}\right) \frac{Q^{6}}{360 \langle M \rangle^{3}}
\nonumber \\
&&\left. + \langle R_{n}^{2}\rangle_\text{eff} \langle R_{n}^{4}\rangle_\text{eff} \sum_{i}\left( X_{i} N_{i}^{2} M_{i}^{4} \right) \left( \frac{T}{E^{2}}  \right) \frac{Q^{6}}{360 \langle M \rangle^{3}} + \cdots \right],
\label{eq:finalform}
\end{eqnarray}
\end{widetext}

\begin{table}[htdp]
\caption{Numerical values for the different effective moments of germanium and
xenon as well as the percent difference between definitions. The definitions
for the effective moments are given in equations \ref{r21}-\ref{r42}. The
values of $(\langle R_{n}^{2} \rangle^{2})_\text{eff}^{1/4}$ are compared to
those of $\langle R^{2}_{n} \rangle_\text{eff,1}^{1/2}$.  }
\begin{center}
\begin{tabular}{|ccc|ccc|ccc|}
\hline
\hline
  & & & & Ge & & & Xe &  \\
 \hline
  & $\langle R^{2}_{n} \rangle_\text{eff,1}^{1/2}$ (fm) & & & 4.0495  & & & 4.8664 &  \\
  & $\langle R^{2}_{n} \rangle_\text{eff,2}^{1/2}$ (fm) & & & 4.0505 & & & 4.8668 &  \\
 & \% Difference & & & 0.02 & & & 0.009 &  \\
   & & & &  & & &  &  \\
 & $\langle R^{4}_{n} \rangle_\text{eff,1}^{1/4}$ (fm) & & & 4.3765 & & &  5.2064  & \\
 & $\langle R^{4}_{n} \rangle_\text{eff,2}^{1/4}$ (fm) & & & 4.3774 & & &  5.2068  & \\
 & \% Difference & & & 0.02 & & & 0.009 &  \\
  & & & &  & & &  &  \\
  & $(\langle R_{n}^{2} \rangle^{2})_\text{eff}^{1/4}$ (fm) & & &  4.0509 & & & 4.8670 & \\
  & \% Difference & & & 0.001 & & & 0.01 &  \\
\hline
\hline
\end{tabular}
\end{center}
\label{table:effMomComparison}
\end{table}%

The approximations leading to Eq. (\ref{eq:finalform}) from Eq. (\ref{eq:complicated-expression}) 
cause an error of ~0.01\% over the entire event curve.

We define effective moments for the protons in the same way as for neutrons.
The terms in the cross section that involve both neutron moments and proton
moments can be calculated from these effective moments.  Since the proton
moments are known quite accurately from electron scattering, we are finally
able to represent the recoil distribution in terms of just two unknown
parameters, $\langle R_{n}^{2} \rangle_\text{eff}$ and $\langle R_{n}^{4}
\rangle_\text{eff}$, or three for xenon.

\subsection{Density Functional Theory Calculations of Moments}

Our Monte-Carlo simulations require the knowledge of the radii and moments
$\langle R_{n,p}^{k}\rangle$ of the neutron and proton distributions in both
even-even and odd-even isotopes. In this work, we compute these
quantities in DFT with Skyrme functionals.  We use nine common
parameterizations of the Skyrme functional (SkM* \cite{Bar82}, SkI3
\cite{Reinhard:1995}, SLy4 \cite{Chabanat:1998}, SLy5 \cite{Chabanat:1998}, SkX
\cite{Brown:1998},HFB9 \cite{Goriely:1999}, SkO \cite{Reinhard:1999}, UNEDF0
\cite{UNEDF0} and UNEDF1 \cite{UNEDF1}).  These functionals are characterized
by relatively different nuclear matter, surface, and deformation properties,
and therefore provide a good ``statistical'' sample of Skyrme functionals. 

We model pairing correlations with a density-dependent delta pairing force with
mixed volume-surface characteristics and the Lipkin-Nogami prescription to
approximate particle number projection. For each element, we fit the strength
of the pairing force to odd-even mass differences according to the general
procedure outlined in \cite{UNEDF0}: in $^{40}$Ar for argon, $^{72}$Ge for
germanium and $^{130}$Xe for xenon. Only parameterizations UNEDF0 and UNEDF1
are accompanied by specific prescriptions for the pairing channel.  We compute
the ground-state in odd-mass isotopes by performing systematic blocking
calculations: for a given odd-mass isotope, we consider all blocking
configurations within 2 MeV of the ground state of the neighboring even-even
nucleus, and take the ground state of the odd isotope to be the configuration
yielding the lowest energy with the correct spin. 

We carry out all these calculations with the latest version of the DFT solver
HFBTHO \cite{HFBTHOnew}. The code solves the Skyrme Hartree-Fock-Bogoliubov
equations in a harmonic oscillator basis with axial symmetry and, therefore,
time-reversal invariance.  We perform all calculations in a full HO basis of 20
shells with a basis frequency parameter defined by $\hbar\omega = 1.2 \times
41/A^{1/3}$. In argon and germanium isotopes the basis is spherical, while in
xenon isotopes, the ground states of which are weakly deformed, its deformation
is $\beta_2 =0.3$. Such characteristics ensure excellent convergence of the
results \cite{Nikolov:2011}. We do the blocking calculations in the equal
filling approximation \cite{PerezMartin:2008}, which agrees with the full
blocking prescription to within 100 keV or less \cite{Schunck:2010}.

%
%

\section{Results and Discussion}
\label{sec:simulations}

\subsection{Monte-Carlo simulations}
We use a simple Monte Carlo simulation to give an idea of how accurately the
nuclear moments can be determined. We assume that a detector, filled with
either $^{40}$Ar, natural germanium, or natural xenon, experiences a flux from
the decay of pions at rest of $3 \times 10^{7}$ neutrinos/cm$^{2}$/sec in each
flavor for one year. The neutrino production rate at the Spallation Neutron
Source at Oak Ridge National Laboratory \cite{Scholberg:2005qs},
DAE$\delta$ALUS \cite{Conrad:2010eu,Conrad:2009mh} and the European Spallation
Source \cite{Drexlin1998193} ranges from $1\times 10^{15}$ neutrinos/sec to
$3.5\times 10^{15}$ neutrinos/sec of each flavor.  A flux of $3 \times 10^{7}$
neutrinos/cm$^{2}$/sec corresponds to detectors placed approximately 16 m from
the source at SNS, 18 m from the source at DAE$\delta$ALUS, and 30 m from the
source at ESS.

We perform a Monte Carlo simulation that includes statistical error and
uncertainty on the beam normalization.  Another significant source of
systematic error is the uncertainty in the detection efficiency.  We discuss
this in Sec.\ III B, and for the purposes of the Monte Carlo assume 100\%
detection efficiency.  We also assume that leptons and photons produced by
charge-current and inelastic neutral-current scattering can be detected and the
corresponding events efficiently rejected as background.

We calculate a recoil curve assuming that the true nuclear density
distributions are given by the Skyrme model SkM$^{*}$.  We place the events
into bins based on energy, and then add Gaussian-distributed statistical noise
to each bin.  We then take the general form factor from Eq.\ \ref{eq:finalform}
and use $\chi^{2}$ minimization to find the optimal values of $\langle
R^{2}_{n} \rangle$, $\langle R^{4}_{n} \rangle$, or the effective moments, and
the beam normalization.  In the case of xenon, we use the same procedure to
find the optimal value of $\langle R^{6}_{n} \rangle$.  Typical values of
$\chi^{2}$ range from 0.5 to 10.

We do separate sets of runs, some assuming that the normalization of the flux
is determined by other means, and some allowing for an uncertainty of $\pm10\%$
in the normalization. Because background is anticipated to be substantial at
high and low energies, we exclude the highest and lowest bins from the
$\chi^{2}$ minimization. The energy ranges included are 5-120 keV for
$^{40}$Ar, 5-70 keV for Ge, and 5-40 keV for Xe. All energy bins are 10 keV
wide except for the lowest, which is 5 keV wide. Finally, we assign confidence
levels to closed areas on an $\langle R^{2}_{n} \rangle$ vs.\ $\langle
R^{4}_{n} \rangle$ plot by running the Monte Carlo many times and dividing the
number of times the minimum-$\chi^2$ result falls in that area by the total
number of runs.

With this setup, we vary the size of the detector until the $\langle R^{2}_{n}
\rangle^{1/2}$ or $\langle R^{2}_{n} \rangle_\text{eff}^{1/2}$ inside the 91\%
confidence region vary by only about $\pm 5 \%$ from the best values.  For that
level of precision, 3.5 tonnes of argon are necessary, 1.5 tonnes of germanium,
and 300 kg of xenon. The required detector mass decreases with atomic size
because the event rate increases roughly as $N^2$. 

The sizes of current and proposed cryogenic detectors can give an idea of the
feasibility of this measurement.  In the case of argon, the existing ICARUS
T600 detector contains 500 tons of LAr \cite{Amerio2004329}, suggesting that a
detector big enough to for measure the form factor is feasible.  For germanium,
existing dark-matter detectors such as CDMS II \cite{CDMS:2010} consist of a few kg of germanium.  The TEXONO-CDEX program is currently using a 1 kg high-purity germanium detector for neutrino physics and dark matter searches \cite{texono:2011}.  The MAJORANA \cite{majorana:2011} and GERDA \cite{gerda:2004} double-beta decay experiments will soon deploy about 40 kg of
germanium enriched in $^{76}$Ge.  One proposed experiment, GEODM
\cite{Anderson:2011bi,CDMS:2010}, would be made up of 300 $\sim$5 kg Ge
crystals, making a total mass of $\sim$ 1.5 tonnes.  Existing xenon detectors,
such as XENON100 \cite{Aprile:2012} and LUX \cite{McKinsey:2010}, are made up
of on the order of a few hundred kg of xenon, approximately the amount required
for a form factor measurement.  A proposed experiment, the LUX-ZEPLIN project,
will use 1.5 tonnes of Xe \cite{McKinsey:2010}. 

\begin{figure}[h]
\includegraphics[width=\linewidth]{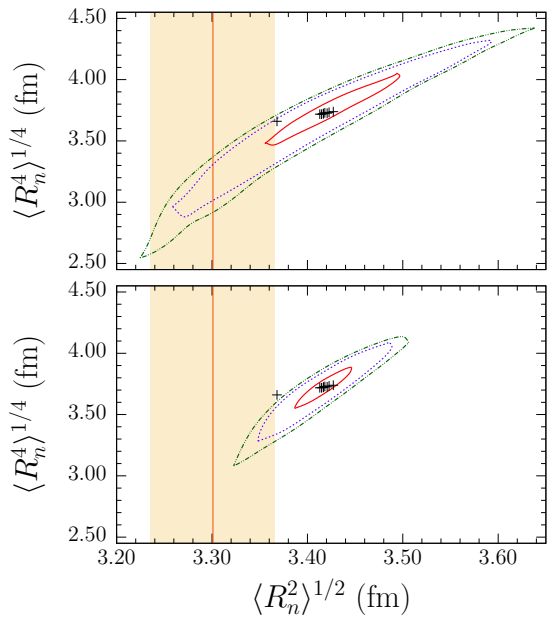} 
\caption{Confidence regions in the $\langle R^{2}_{n} \rangle^{1/2}$--
$\langle R^{4}_{n}\rangle^{1/4}$ plane for an argon detector of mass 3.5
tonnes. The curves enclose confidence regions of 40\%, 91\%, and 97\%. The colored vertical band shows the experimental
result reported for the RMS radius, obtained from argon-carbon scattering, in
Ref.\ \cite{Ozawa200260}, and the black crosses are the 
predictions of some commonly used Skyrme functionals, including the functional 
SkM$^*$ that we use to generate the ``data.'' Numerical results are summarized 
in tables \ref{table:MCresults} and \ref{table:MCresultsLnuConstant}. Top panel: 
the neutrino flux is allowed to vary by $\pm 10$ \%. Bottom panel: the flux is 
assumed to be known exactly.}
\label{fig:argonResult} 
\end{figure}

\begin{figure}[h]
\includegraphics[width=\linewidth]{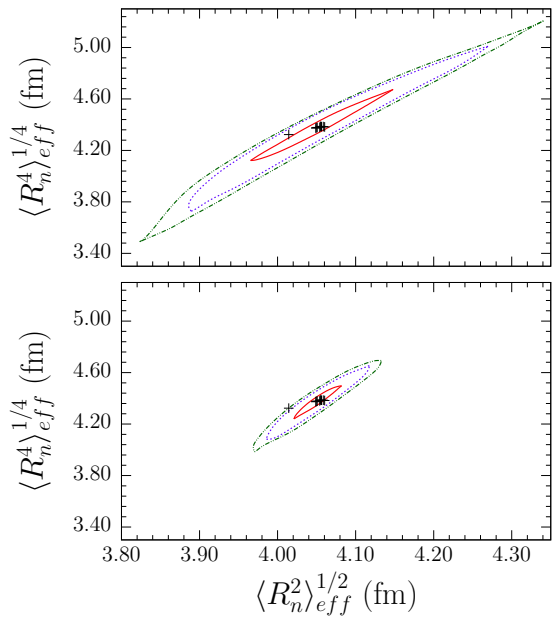}
\caption{Same as Fig.\ \ref{fig:argonResult}, but for effective moments in
germanium, and without experimental result.} 
\label{fig:germaniumResult}
\end{figure}

\begin{figure}[b!]
\includegraphics[width=\linewidth]{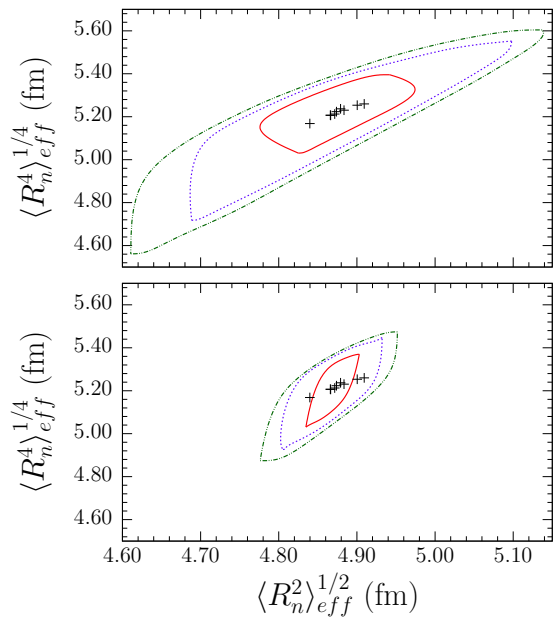}
\caption{Same as Fig.\ \ref{fig:germaniumResult}, but in xenon.} 
\label{fig:xenonResult}
\end{figure}

The results of the analysis appear in Figs.\ \ref{fig:argonResult},
\ref{fig:germaniumResult}, and \ref{fig:xenonResult}. The closed curves
correspond to 40\% confidence, 91\% confidence, and 97\% confidence. As
mentioned above, we considered two cases: one in which the
normalization of the flux is allowed to vary (by $\pm 10 \%$), and a second in
which the normalization is kept constant. The top panels of each figure show
the results with the flux unconstrained within that $10 \%$ range, and the
bottom panels show the same results with the assumption that the flux is known
perfectly.  The colored vertical band in Fig.\ \ref{fig:argonResult} shows a
model-dependent experimental result for the RMS radius, obtained from
argon-carbon scattering in Ref.\ \cite{Ozawa200260}. There is a clear
discrepancy between that result and the predictions of the 9 Skyrme functionals
selected for this study, labeled by small crosses in Fig.\
\ref{fig:argonResult} (the outlyer in Figures
\ref{fig:argonResult}-\ref{fig:germaniumResult} corresponds to the SkX
functional of \cite{Brown:1998}, which predicts systematically smaller radii
than other functionals). Those functionals include SkM$^*$, the one we use to
generate the ``data.'' This discrepancy is mentioned by Ozawa {\it et al.}, but
no explanation is offered.  While we marginalize over $\langle R_{n}^{6}\rangle_\text{eff}$ for xenon, the quantity is poorly constrained and
not included in the plot in Fig.\ \ref{fig:xenonResult}.  Numerical results at
the 91\% confidence level for the mean, minimum, and maximum of the (effective)
RMS neutron radius and fourth moment, (and sixth moment in xenon) appear in
Tabs.\ \ref{table:MCresults} and \ref{table:MCresultsLnuConstant}.

\begin{table*}
\caption{Numerical results at the 91\% confidence level for the 3.5 tonne 
$^{40}$Ar detector, the 1.5 tones Ge detector, and the 300 kg Xe detector with 
$L_{\nu}$ allowed to vary by $\pm 10\%$.  The first column contains the element, 
the second the moment or effective moment considered in the corresponding row, 
the third the calculated values of the moments or effective moments for the 
Skyrme model SkM$^*$, the fourth the mean values for the moments or effective 
moments, produced by the Monte Carlo, the fifth the percent difference between 
the mean values and the SkM$^*$ values, the sixth the minimum values chosen by 
the Monte Carlo, and the seventh the percent difference between the minimum and 
the mean value. The eighth column gives the maximum values chosen by the Monte 
Carlo, and the ninth column gives the percent difference between the maximum 
and the mean values. } 
\label{table:MCresults}
\begin{center}
\begin{ruledtabular}
\begin{tabular}{|c|c|c|c|>{\centering}m{2.2cm}|c|>{\centering}m{2.2cm}|c|>{\centering}m{2.2cm}|}
 & & SkM$^*$ values & Mean & \% Difference (from SkM$^*$) & Min & \% Difference
 (from mean)  & Max  & \% Difference (from mean)\tabularnewline
\hline
\multirow{2}{*}{$^{40}$Ar} & $\langle R_{n}^{2} \rangle^{1/2}$ (fm) & 3.4168 &
3.4103 & -0.2 & 3.2587 & -4  & 3.5999 & +6 \tabularnewline
 & $\langle R_{n}^{4}\rangle^{1/4}$ (fm) & 3.7233 & 3.6576 & -2 &
2.8304 & -23 & 4.3210 & +18\tabularnewline
\hline
 \multirow{2}{*}{Ge} & $\langle R_{n}^{2} \rangle_\text{eff}^{1/2}$ (fm) &
 4.0495 & 4.0516 & +0.05 & 3.8792 & -4  & 4.2697 & +5\tabularnewline
  & $\langle R_{n}^{4}\rangle_\text{eff}^{1/4}$ (fm) & 4.3765 &
4.3603 & -0.4 & 3.7276 & -15  & 5.0096 & +15 \tabularnewline
 \hline
  \multirow{3}{*}{Xe} & $\langle R_{n}^{2} \rangle_\text{eff}^{1/2}$ (fm) &
  4.8664 & 4.8648 & -0.001 & 4.6788 & -4  & 5.0980 & +5\tabularnewline
 & $\langle R_{n}^{4}\rangle_\text{eff}^{1/4}$ (fm) & 5.2064 &
5.1914 & -0.3 & 4.7180 & -10  & 5.5521 & +7 \tabularnewline
 & $\langle R_{n}^{6}\rangle_\text{eff}^{1/6}$ (fm) & 5.4887 &
5.3149 & -3 & 0.5491 & -90  & 10.433 & +97 \tabularnewline
 \end{tabular}
 \end{ruledtabular}
 \end{center}
 \end{table*}

\begin{table*}
\caption{Same as table \ref{table:MCresults}, except for $L_{\nu}$ held constant. } 
\label{table:MCresultsLnuConstant}
\begin{center}
\begin{ruledtabular}
\begin{tabular}{|c|c|c|c|>{\centering}m{2.2cm}|c|>{\centering}m{2.2cm}|c|>{\centering}m{2.2cm}|}
 & & SkM$^*$ values & Mean  & \% Difference (from SkM$^*$) & Min  & \%
 Difference (from mean)  & Max  & \% Difference (from mean)\tabularnewline
\hline
\multirow{2}{*}{$^{40}$Ar} & $\langle R_{n}^{2} \rangle^{1/2}$ (fm)  & 3.4168 &
3.4154 & -0.04 & 3.3483 & -2  & 3.4933 & +2 \tabularnewline
  & $\langle R_{n}^{4}\rangle^{1/4}$ (fm) & 3.7233 & 3.7018 & -0.6 &
3.2826 & -11 & 4.0865 & +10 \tabularnewline
 \hline
 \multirow{2}{*}{Ge} & $\langle R_{n}^{2} \rangle_\text{eff}^{1/2}$ (fm) & 4.0495 &
 4.0491 & -0.009 & 3.9857 & -1.6  & 4.1175 & +1.7\tabularnewline
  & $\langle R_{n}^{4}\rangle_\text{eff}^{1/4}$ (fm) & 4.3765 & 4.3679 &
-0.2 & 4.0826 & -7  & 4.6546 & +7 \tabularnewline
 \hline
  \multirow{3}{*}{Xe} & $\langle R_{n}^{2} \rangle_\text{eff}^{1/2}$ (fm) & 4.8664 &
  4.8654 & -0.02 & 4.7958 & -1.4  & 4.9323 & +1.4 \tabularnewline
 & $\langle R_{n}^{4}\rangle_\text{eff}^{1/4}$ (fm) & 5.2064 & 5.1990 &
-0.14 & 4.9265 & -5  & 5.4478 & +5 \tabularnewline
  & $\langle R_{n}^{6}\rangle_\text{eff}^{1/6}$ (fm) & 5.4887 &
5.3877 & -1.8 & 0.5491 & -90  & 10.433 & +94 \tabularnewline
 \end{tabular}
 \end{ruledtabular}
 \end{center}
 \end{table*}

\subsection{Discussion}

There are several noticeable differences among figures \ref{fig:argonResult},
\ref{fig:germaniumResult}, and \ref{fig:xenonResult}.  The first is the dependence on
$\langle R_{n}^{4} \rangle_\text{eff}^{1/4}$.  For $^{40}$Ar, the range of
plausible values is large compared to the range of $\langle R_{n}^{2}
\rangle^{1/2}$.  As the nuclei grow in size to germanium and then xenon, the
range of $\langle R_{n}^{4} \rangle_\text{eff}^{1/4}$ gets smaller.  In germanium
and xenon, it is comparable to the range of the effective RMS neutron radius.

We can explain this difference by isolating the effects of the (effective)
fourth moment on the recoil distributions.  In $^{40}$Ar, a 10\% change in the
fourth moment results in approximately 0.2\% more events, as compared to the
1.2\% difference with a 10\% change in the RMS radius.  In comparison, the same
change in $\langle R_{n}^{4} \rangle_\text{eff}^{1/4}$ of 10\% results in 1.3\%
and 3\% more events for germanium and xenon, respectively, as compared to 6\%
and 8\% from a 10\% change in $\langle R_{n}^{2} \rangle_\text{eff}^{1/2}$. The
nuclear recoil energy at which this difference is concentrated decreases and
the importance of the fourth moment relative to the RMS radius increases as the
nuclear mass increases.  

Our ability to learn about the nuclear quantities $\langle R_{n}^{2}
\rangle_\text{eff}^{1/2}$ and $\langle R_{n}^{4} \rangle_\text{eff}^{1/4}$
obviously improves if we can get an independent handle on the beam
normalization.  This can be seen clearly by comparing the top and bottom panels
in Figs. 3, 4 and 5.  The range in $\langle R_{n}^{2} \rangle_\text{eff}^{1/2}$
in all three elements shrinks to $\pm 2 \%$ at the 91\% level in the bottom
panels, where the beam normalization is known exactly.  Likewise, the range in
$\langle R_{n}^{4} \rangle_\text{eff}^{1/4}$ decreases for all three elements.
The effect is most dramatic for $^{40}$Ar, where the uncertainty decreases to
$\sim \pm 10\%$, but it exists in both germanium and xenon as well.

As mentioned above, we consider the systematic error of the uncertainty in
detection efficiency.  In order to study the effect, we remove the statistical
error and randomly distribute systematic errors, proportional to the number of
events in the bin, at the level of 10\%, 1\% and 0.1\% in each bin in an
uncorrelated manner.  At the 10\% level, detectors of the size considered here
lose the ability to make any useful measurement of the neutron radius.  When we
lower the uncertainty in detection efficiency to 1\%, measurements of the
neutron radius to $\sim \pm$5-7\% are possible.  At this level, the range in
the value of the fourth moment is $\sim \pm$20\%.  If we lower the uncertainty
in detector efficiency to 0.1\%, the neutron radius can be measured to better
than $\sim \pm$1\%, and the fourth moment to $\pm$2\%.

%
%

\section{Conclusions}
Neutron radii are not only of fundamental interest for nuclear structure but
are also needed to fully analyze supernova-neutrino signals
\cite{Horowitz:2003cz} and interpret measurements of the Weinberg angle or of
neutrino magnetic moments \cite{Scholberg:2005qs}. At present the distributions
of neutrons in nuclei is not known nearly as well as those of protons.

We have presented a model-independent method, involving the Taylor expansion of
the scattering form factor, for extracting the RMS radius and fourth moment of
the neutron density distribution in certain nuclei from the nuclear-recoil
distribution in a neutrino-scattering experiment. The radius and fourth moment
can also be be calculated theoretically, so that our technique will provide a
straightforward connection between theory and experiment. To obtain a rough
estimate of the effectiveness of this approach, we considered a stopped pion
neutrino source of $3\times 10^{7}$ neutrinos/cm$^{2}$/s and liquid Ar, Ge, and
Xe detectors in the tonne range. We conclude that it is possible to
determine the neutron radius to a few percent if the uncorrelated error on the
efficiency is less than 1\%.  The detailed analysis of the shape of the
recoil spectrum in a cryogenic detector, such as the one we have suggested
here, has not previously been considered.  More detailed simulations of
realistic experimental setups are therefore required for definitive feasibility
studies.

%
%

\section{Acknowledgements}
We thank Kate Scholberg for helpful discussions.  This work was supported by a
GAANN fellowship (KP), by DOE contract DE-FG02-02ER41216 (KP and GM), by
DOE Contract DE-FG02-97ER41019 (JE). It was partly performed under the auspices 
of the US Department of Energy by the Lawrence Livermore National Laboratory 
under Contract DE-AC52-07NA27344, and funding was also provided by the United 
States Department of Energy Office of Science, Nuclear Physics Program pursuant 
to Contract DE-AC52-07NA27344 Clause B-9999, Clause H-9999 and the American 
Recovery and Reinvestment Act, Pub. L. 111-5 (NS).

%
%

\bibliography{bibliography}

\end{document}